\documentclass{article}
\usepackage{authblk,xcolor,fancyvrb,fullpage}

\newcommand{\occ}{\ensuremath{\mathrm{occ}}}
\newcommand{\PBWT}{\ensuremath{\mathrm{PBWT}}}
\newcommand{\rank}{\ensuremath{\mathrm{rank}}}

\begin{document}

\title{Teaching the Burrows-Wheeler Transform via\\the Positional Burrows-Wheeler Transform}
\author[1]{Travis Gagie}
\affil[1]{Dalhousie University, Canada}
\author[2]{Giovanni Manzini}
\affil[2]{University of Pisa, Italy}
\author[3]{Marinella Sciortino}
\affil[3]{University of Palermo, Italy}
\date{}
\maketitle

\begin{abstract}
The Burrows-Wheeler Transform (BWT) is often taught in undergraduate courses on algorithmic bioinformatics, because it underlies the FM-index and thus important tools such as Bowtie and BWA.  Its admirers consider the BWT a thing of beauty but, despite thousands of pages being written about it over nearly thirty years, to undergraduates seeing it for the first time it still often seems like magic.  Some who persevere are later shown the Positional BWT (PBWT), which was published twenty years after the BWT.  In this paper we argue that the PBWT should be taught {\em before} the BWT.

We first use the PBWT's close relation to a right-to-left radix sort to explain how to use it as a fast and space-efficient index for {\em positional search} on a set of strings (that is, given a pattern and a position, quickly list the strings containing that pattern starting in that position).  We then observe that {\em prefix search} (listing all the strings that start with the pattern) is an easy special case of positional search, and that prefix search on the suffixes of a single string is equivalent to {\em substring search} in that string (listing all the starting positions of occurrences of the pattern in the string).

Storing na\"ively a PBWT of the suffixes of a string is space-{\em inefficient} but, in even reasonably small examples, most of its columns are nearly the same.  It is not difficult to show that if we store a PBWT of the cyclic shifts of the string, instead of its suffixes, then all the columns are exactly the same --- and equal to the BWT of the string.  Thus we can teach the BWT and the FM-index via the PBWT.
\end{abstract}

\section{Introduction}
\label{sec:introduction}

The Burrows-Wheeler Transform (BWT)~\cite{burrows1994block} was published nearly thirty years ago, initially as a pre-processing step to ease data compression.  Although it is intuitively clear why in practice a transformed string is more compressible --- the transform sorts the characters in the string into the lexicographic order of the suffixes that immediately follow them in the string, and thus moves together characters with similar contexts~\cite{fenwick1996burrows,manzini2001analysis,nelson1996data} --- most students do not immediately see why it is invertible.  Although its admirers consider it a thing of beauty and despite thousands of pages being written about the BWT (see, e.g.,~\cite{adjeroh2008burrows} and references therein), to undergraduates seeing if for the first time it still often seems like magic.

If the BWT were used only in data compression, difficulties teaching it could probably be ignored, but it underlies the FM-index~\cite{ferragina2005indexing,willets2003full} and thus plays a key role in important tools in bioinformatics such as the popular DNA aligners Bowtie~\cite{langmead2009ultrafast,langmead2012fast} and BWA~\cite{li2009fast}.  It is therefore often included in undergraduate courses on algorithmic bioinformatics (see, e.g., {\tt https://www.coursera.org/learn/dna-sequencing}), where it remains one of the more challenging data structures to teach.

We conjecture that the main reason students struggle to grasp the BWT and the FM-index the first time they see them --- and often the second and third time as well --- is because it is difficult to break the standard presentation down into easily-understood steps that a good student should be able to figure out themselves with some guidance.  Either a student sees why the BWT is invertible, or they do not, and either they understand how backward search works, or they do not; there is not much middle ground.

In this paper we propose a new way to present the BWT and FM-index, starting from the positional BWT (PBWT), which Durbin~\cite{durbin2014efficient} developed from the BWT.  Probably because the PBWT was developed twenty years after the BWT and Durbin presented it as a data structure for the relatively advanced task of haplotype matching, the PBWT is usually taught to students who have already seen the BWT and persevered.  We argue that the PBWT should be taught {\em before} the BWT.

Unusually, we do not start by discussing how to invert the BWT or PBWT, because now most BWTs are built and used in bioinformatics indexes without ever being inverted.  We also do not consider compression because, for example, Bowtie and BWA do not try to achieve a space bound better than $n \lg \sigma$, where $n$ is the length of the indexed string and $\sigma$ is the size of its alphabet (so 4 for DNA).  Instead, we start in Section~\ref{sec:positional} by considering {\em positional search} on a set of strings --- given a pattern and a position, quickly list the strings containing that pattern starting in that position --- and there and in Section~\ref{sec:saving} use the PBWT's close relation to a right-to-left radix sort to explain how to use it as a fast and space-efficient index for this problem.  (Admittedly, Durbin defined the PBWT in terms of co-lexicographic sorting and we define it in terms of lexicographic sorting, but the definitions are symmetric.)  We believe that good undergraduate students should be able to follow this development easily and, more importantly, be guided to discover it for themselves.

We then observe in Section~\ref{sec:substring} that {\em prefix search} --- listing all the strings that start with the pattern --- is an easy special case of positional search, and that prefix search on the suffixes of a single string is equivalent to {\em substring search} in that string --- listing all the starting positions of occurrences of the pattern in the string.  Of course, storing na\"ively a PBWT of the suffixes of a string is space-{\em inefficient} --- it takes space quadratic in the length of the string --- but, in even reasonably small examples, most of its columns are nearly the same.  It is not difficult to show that if we store a PBWT of the cyclic shifts of the string, instead of its suffixes, then all the columns are exactly the same --- and equal to the BWT of the string.  We then describe how the suffix-array sample used in the FM-index can be developed from the PBWT.  Thus we can teach the BWT and the FM-index via the PBWT.

\section{Positional Search}
\label{sec:positional}

Suppose we are given $n$ strings $S_0, \ldots, S_{n - 1}$ and asked to store them such that we can support fast {\em positional search}: that is, such that later, given a pattern $P [0..m - 1]$ and a position $k$, we can quickly list the indexes of the strings in which $P$ occurs starting in position $k$.  For the sake of simplicity, let us assume each string has length $n$.  For example, if the strings are the ones shown in Figure~\ref{fig:strings}, $P = \mathtt{AGA}$ and $k = 3$, then we should return 1, 4 and 5.

\begin{figure}[t!]
\begin{eqnarray*}
S_0 & = & \mathtt{GATTACAT} \\
S_1 & = & \mathtt{TAGAGATA} \\
S_2 & = & \mathtt{CATCACAT} \\
S_3 & = & \mathtt{TACATACA} \\
S_4 & = & \mathtt{GATAGATA} \\
S_5 & = & \mathtt{TAAAGAGC} \\
S_6 & = & \mathtt{ATTACCAT} \\
S_7 & = & \mathtt{ACATTACT}
\end{eqnarray*}
\caption{A collection of $n = 8$ strings of length $n = 8$ over an alphabet of size $\sigma = 4$.}
\label{fig:strings}
\end{figure}

A fairly obvious solution to this problem is to store, for $0 \leq j \leq n - 1$, a permutation $\pi_j$ on $\{0, \ldots, n - 1\}$ such that $S_{\pi_j (i)} [j..n - 1]$ is lexicographically $i$th --- we count from 0 --- among the the strings' suffixes $S_0 [j..n - 1], \ldots, S_{n - 1} [j..n - 1]$ starting in position $j$.  Given $P$ and $k$, we can use $\pi_k$ to perform binary searches on the suffixes $S_{\pi_k (0)} [k..n - 1], \ldots, S_{\pi_k (n - 1)} [k..n - 1]$ starting in position $k$ to find the lexicographic ranks $f_k$ and $\ell_k$ of the first and last such suffixes starting with $P$, in time $O (m \log n)$.  We can then report $\pi_k (f_k), \ldots, \pi_k (\ell_k)$ in $O (\occ_k)$ time, where $\occ_k = \ell_k - f_k + 1$ is the number of strings containing $P$ starting in position $k$.

A fairly obvious way to compute $\pi_0, \ldots, \pi_{n - 1}$ quickly is to treat the strings as $n$-digit numbers in base $\sigma$ --- the size of the alphabet --- and perform a right-to-left radix sort on them.  Figure~\ref{fig:sort} shows {\tt C} code for computing and printing $\pi_1, \ldots, \pi_n$ as the columns of a matrix.  The left side of the figure is straightforward, so we discuss only the {\tt main()} procedure.

\begin{figure}[t!]
\begin{center}
\begin{BVerbatim}
#include <stdio.h>

#define N       8
#define SIGMA   4

char S[N][N] = {"GATTACAT",
                "TAGAGATA",
                "CATCACAT",
                "TACATACA",
                "GATAGATA",
                "TAAAGAGC",
                "ATTACCAT",
                "ACATTACT"};

int Pi[N][N + 1];

int alphRank(char c) {
  switch (c) {
    case 'A':
      return(0);
    case 'C':
      return(1);
    case 'G':
      return(2);
    case 'T':
      return(3);
  }
}

void printPi() {

  for (int i = 0; i < N; i++) {
    for (int j = 0; j < N; j++) {
      printf("%i\t", Pi[i][j]);
    }
    printf("\n");
  }
}
\end{BVerbatim}
\hspace{5ex}
\rule{.1ex}{105ex}
\hspace{5ex}
\begin{BVerbatim}
void main() {
  
  for (int i = 0; i < N; i++) {
    Pi[i][N] = i;
  }
  
  for (int j = N - 1; j >= 0; j--) {
    
    int Freq[SIGMA] = {0};
    
    for (int i = 0; i < N; i++) {
      Freq[alphRank(S[i][j])]++;
    }
    
    int C[SIGMA];
    
    C[0] = 0;
    
    for (int a = 1; a < SIGMA; a++) {
      C[a] = C[a - 1] + Freq[a - 1];
    }
        
    int Rank[SIGMA] = {0};
    
    for (int i = 0; i < N; i++) {
      int a = alphRank(S[Pi[i][j + 1]][j]);
      Pi[C[a] + Rank[a]][j] = Pi[i][j + 1];
      Rank[a]++;
    }
  }
  
  printPi();
}





\end{BVerbatim}
\caption{{\tt C} code for performing a right-to-left radix sort of the strings in our example, and printing the permutations $\pi_0, \ldots, \pi_{n - 1}$ as the columns of a matrix.}
\label{fig:sort}
\end{center}
\end{figure}

We set $\pi_n = 0, \ldots, n - 1$ for convenience --- as a lexicographic order of the empty strings $S_0 [n..n - 1], \ldots, S_{n - 1} [n..n - 1]$ --- and then compute $\pi_j$ for for the values of $j$ from $n - 1$ to 0 in turn.  For each value of $j$, we first compute the frequency of each distinct character in $S_0 [j]\,S_1 [j] \ldots S_{n - 1} [j]$ and then compute, for $0 \leq a \leq \sigma - 1$, the number $C [a]$ of characters in $S_0 [j]\,S_1 [j] \ldots S_{n - 1} [j]$ that are lexicographically strictly less than the $a$th character in the alphabet.  The name $C$ of the array is admittedly not very descriptive, but we use it for consistency with other papers about the BWT and FM-index.

Notice that $S_i [j] \prec S_{i'} [j]$ implies $S_i [j..n - 1] \prec S_{i'} [j..n - 1]$, so the values of $i$ for which $S_i [j]$ is the $a$th character in the alphabet are, in some order, the $C_j [a]$th through $(C_j [a + 1] - 1)$st values in $\pi_j$.  Moreover, $S_i [j] = S_{i'} [j]$ and $S_i [j + 1..n - 1] \prec S_{i'} [j + 1..n - 1]$ together imply $S_i [j..n - 1] \prec S_{i'} [j..n - 1]$.

It follows that, since we already have the lexicographic order $\pi_{j + 1}$ of $S_0 [j + 1..n - 1], \ldots, S_{n - 1} [j + 1..n - 1]$, we can efficiently compute $\pi_j$ by scanning the string $S_{\pi_{j + 1} (0)} [j]\,S_{\pi_{j + 1} (1)} [j] \ldots S_{\pi_{j + 1} (n - 1)} [j]$.  If the $i$th character in that string is the $a$th character in the alphabet, then we set
\[\pi_j \left( \rule{0ex}{2ex} C [a] + \rank_a (\pi_{j + 1} (i)) \right) = \pi_{j + 1} (i)\,,\]
where $\rank_a (\pi_{j + 1} (i))$ is the number of copies of the $a$th character in the alphabet that we have already seen.

The lines
\begin{center}
\begin{BVerbatim}
int Rank[SIGMA] = {0};

for (int i = 0; i < N; i++) {
  int a = alphRank(S[Pi[i][j + 1]][j]);
  Pi[C[a] + Rank[a]][j] = Pi[i][j + 1];
  Rank[a]++;
}
\end{BVerbatim}
\end{center}
perform this step in the radix sort.

Although the preceding paragraph may seem quite technical, we emphasize that it is only a detailed description of a right-to-left radix sort on a set of strings, which should not be out of place in an undergraduate class.  Figure~\ref{fig:output} shows the output of the code in Figure~\ref{fig:sort}.

\begin{figure}[t!]
\begin{center}
\begin{BVerbatim}
7       5       5       6       0       3       0       1
6       3       7       5       2       7       2       3
2       1       3       1       6       5       6       4
4       4       1       4       5       1       3       5
0       2       6       3       1       4       7       0
5       0       4       2       4       0       5       2
3       7       2       0       3       2       1       6
1       6       0       7       7       6       4       7
\end{BVerbatim}
\caption{The output of the code in Figure~\ref{fig:sort}, showing the permutations $\pi_0, \ldots, \pi_{n - 1}$ for our example as the columns of a matrix.}
\label{fig:output}
\end{center}
\end{figure}

\section{Saving Space}
\label{sec:saving}

Storing $S_0, \ldots, S_{n - 1}$ takes $O (n^2 \log \sigma)$ bits, but storing na\"ively $\pi_0, \ldots, \pi_{n - 1}$ as well takes $\Omega (n^2 \log n)$ bits in the worst case.  A fairly obvious question is whether we can store $S_0, \ldots, S_{n - 1}$ and $\pi_0, \ldots, \pi_{n - 1}$ in a total of $O (n^2 \log \sigma)$ bits and still support fast positional searches.

Of course, if we do not care about the speed of the positional searches, then we can simply store $S_0, \ldots, S_{n - 1}$ in $O (n^2 \log \sigma)$ bits and, for each query $P$ and $k$, build $\pi_{n - 1}, \ldots, \pi_k$ column by column --- discarding each column when we are done with it to save space --- in $O (n^2 - n k)$ time.  After we build $\pi_k$, we can perform binary searches as before and thus list the indexes of the strings in which $P$ occurs starting in position $k$ in $O (n^2 - n k + m \log n + \occ_k)$ total time.

As a fairly obvious trade-off between these two extremes, we can store $\pi_j$ for every $\lceil \lg n \rceil$th value of $j$ in $O (n^2)$ bits and, when given $P$ and $k$, start with $\pi_j$ for the smallest value of $j$ at least $k$ and build back to $\pi_k$, in $O (n \log n)$ time.  This brings our total query time down slightly, to $O (n \log n + m \log n + \occ_k)$.

A slightly less obvious option is to store $\pi_j$ for every $\lceil \lg n \rceil$th value of $j$ and, when given $P$ and $k$, at first consider only the suffix of $P [m + j - k..m]$ that should start at the next column $j$ for which we have $\pi_j$ stored.  We can use that $\pi_j$ to perform binary searches on the suffixes $S_{\pi_j (0)} [j..n - 1], \ldots, S_{\pi_j (n - 1)} [j..n - 1]$ starting in position $j$ to find the lexicographic ranks $f_j$ and $\ell_j$ of the first and last such suffixes of the strings, starting with $P [m + j - k..m]$.

Once we have $f_j$ and $\ell_j$, we can scan the string $S_{\pi_j (0)} [j - 1]\,S_{\pi_j (1)} [j - 1] \ldots S_{\pi_j (n - 1)} [j - 1]$ --- which we can either build or have stored without increasing our space bound --- and compute the lexicographic ranks $f_{j - 1}$ and $\ell_{j - 1}$ of the suffixes of the strings, starting with $P [m + j - k - 1..m]$.  Working backwards like this, we can find $f_k$ and $\ell_k$.

For our example with $P = \mathtt{AGA}$ and $k = 3$, if we do not have $\pi_3$ or $\pi_4$ stored but we do have $\pi_5 = 3, 7, 5, 1, 4, 0, 2, 6$ stored then we find $f_5 = 0$ and $\ell_5 = 4$ via binary searches, meaning strings $S_3, S_7, S_5, S_1, S_4$ contain $P [2] = \mathtt{A}$ in position 5.  We scan the string
\[S_{\pi_5 (0)} [4]\,S_{\pi_5 (1)} [4] \ldots S_{\pi_5 (7)} [4]
= \mathtt{TTGGGAAC}\]
and find 3 copies of {\tt G} in the range $[f_5 = 0..\ell_5 = 4]$, for strings $S_5, S_1, S_4$.  The code inside the loop \verb!for (int j = N - 1; j >= 0; j--) { }! sets $\pi_4 (3), \ldots, \pi_4 (5)$ to those strings indexes, so $f_4 = 3$ and $\ell_4 = 5$.

Examining the code inside that loop, we can see that if we have stored the array $C$ and a data structure supporting rank queries on the string {\tt TTGGGAAC}, then we can compute $f_4$ and $\ell_4$ from $f_5$ and $\ell_5$ with only two rank queries, instead of scanning the string: one rank query tells us the rank for the first copy of {\tt G} in the interval $[0, \ldots, 4]$ and the other tells us the rank for the last copy of {\tt G} in that interval.

This string {\tt TTGGGAAC} is column $\PBWT [0..n - 1][4]$ in the PBWT for our example.  Changing the line \verb!printf("%i\t", Pi[i][j]);! to \verb!printf("%c\t", S[Pi[i][j + 1]][j]);! in the code in Figure~\ref{fig:sort} causes the code to print out the PBWT, shown in Figure~\ref{fig:PBWT}.  (Notice the last column is unchanged from Figure~\ref{fig:strings}.) 
 Explaining how rank data structures work is beyond the scope of this paper, as it is not necessary for students to understand the PBWT and BWT; they only really need to know they exist, occupy a total of $O (n^2 \log \sigma)$ bits for all the columns of the PBWT, and answer queries quickly. 

\begin{figure}[t!]
\begin{center}
\begin{BVerbatim}
T       A       T       T       T       C       T       T
T       C       A       C       T       C       C       A
T       A       G       A       G       C       T       T
G       A       T       A       G       A       G       A
C       T       C       A       G       A       A       A
G       A       T       A       A       A       A       C
A       A       T       A       A       A       A       T
A       A       A       T       C       A       C       T
\end{BVerbatim}
\caption{The PBWT for our example.}
\label{fig:PBWT}
\end{center}
\end{figure}

The final step in our presentation of the PBWT is to observe that, if we have rank data structures for the columns of the PBWT, then we do not need binary search at all.  We can start at column $k + m$ with $f_{k + m} = 0$ and $\ell_{k + m} = n - 1$ and work backwards as described above, until we find $f_k$ and $\ell_k$.  Once we have the interval $\PBWT [f_k..\ell_k][k]$, we can work backwards from each cell $\PBWT [i][k]$ with $f_k \leq i \leq \ell_k$ until we reach the rightmost column $h \leq k$ for which we have $\pi_h$ stored, which tells us from which string $\PBWT [i]$ originates.  This changes our query time to $O (m f (n) + \occ_k f (n))$, where $f (n)$ is the query time of the rank data structures.

For our example, if the rightmost column $h \leq 3$ for which we have $\pi_h$ stored is 2 then, since $k = 3$ and $m = 3$, we would start with $f_6 = 0$ and $\ell_6 = 7$.  The ranks of the first and last copies of {\tt A} in $\PBWT [0..7][5]$ are 0 and 4, so $f_5 = 0$ and $\ell_5 = 4$.  The ranks of the first and last copies of {\tt G} in $\PBWT [0..4][4]$ are 0 and 2 and there are 3 characters lexicographically less than {\tt G} in $\PBWT [0..7][4]$, so $f_4 = 3$ and $\ell_4 = 5$.  The ranks of the first and last copies of {\tt A} in $\PBWT [3..5][3]$ are 1 and 3, so $f_3 = 1$ and $\ell_3 = 3$.

The first character in $\PBWT [1..3][2]$ is $\PBWT [1][2] = \mathtt{A}$ with rank 0, so it comes from string $S_{\pi_2 (0)} = S_5$.  The second character is $\PBWT [2][2] = \mathtt{G}$ with rank 0 and 3 characters lexicographically less than {\tt G} in $\PBWT [0..7][2]$, so it comes from string $S_{\pi_2 (3)} = S_1$.  The third character is $\PBWT [3][2] = \mathtt{T}$ with rank 1 and 4 characters lexicographically less than {\tt T} in $\PBWT [0..7][2]$, so it comes from string $S_{\pi_2 (5)} = S_4$.

\section{Substring Search}
\label{sec:substring}

Assuming students have followed the development of the PBWT, it remains for us to present the BWT and FM-index in terms of the PBWT, and show how to support fast {\em substring searches}.  For substring search, we are given a single string $S$ and asked to index it such that later, given a pattern $P [0..m - 1]$, we can quickly list all that starting positions of the $\occ$ occurrences of $P$ in $S$.  We note as an aside that we can still apply the techniques we present in this section if we are asked to index a set of strings, in which case we obtain the extended BWT (eBWT)~\cite{mantaci2007extension} of the set of strings instead of the BWT of a single string.

A space-{\em inefficient} way to support substring search is via {\em prefix search}: we store treat each suffix of $S$ as a separate string and index that set of strings such that we can list the indexes of the ones starting with $P$.  Of course, prefix search is a special case of positional search --- and if we store $\pi_0$ then our query time becomes $O (m f (n) + \occ)$.

This na\"ive approach is space-inefficient because it takes $\Omega (n^2 \log \sigma)$ bits in the worst case to index a string that fits in $O (n \log \sigma)$ bits.  Looking at the PBWT for a reasonably small example suggests a route to saving space, however: Figure~\ref{fig:suffixes} shows the PBWT for the suffixes of {\tt GATTAGATACAT}, the columns of which are quite similar to their neighbours.

\begin{figure}[t!]
\begin{center}
\begin{BVerbatim}
T       T       T       T       T       T       T       T       T       T       T       T
T       -       -       -       -       -       -       -       -       -       -       -
T       T       -       -       -       -       -       -       -       -       -       -
C       T       T       -       -       -       -       -       -       -       -       -
G       C       T       T       -       -       -       -       -       -       -       -
G       G       C       T       T       -       -       -       -       -       -       -
A       A       G       C       C       T       -       -       -       -       -       -
A       A       A       G       G       C       T       -       -       -       -       -
A       A       A       A       A       G       C       T       -       -       -       -
A       A       A       A       A       A       A       C       C       -       -       -
T       T       A       A       A       A       A       A       A       C       -       -
A       A       T       A       A       A       A       A       A       A       A       -
\end{BVerbatim}
\caption{The PBWT for the suffixes of of {\tt GATTAGATACAT}.}
\label{fig:suffixes}
\end{center}
\end{figure}

The reason neighbouring columns tend to be similar becomes clear if we consider how we produce the column $j$ of the PBWT by sorting the characters in $S [j..n - 1]$ into the lexicographic order of the suffixes that immediately follow them.  Figure~\ref{fig:comparison} shows the characters in $S [0..11] = \mathtt{GATTAGATACAT}$ and the suffixes we sort them by to get column 0 in Figure~\ref{fig:suffixes}, and the characters in $S [1..11]$ and the suffixes we sort them by to get column 1.  Looking at that figure, it is natural to wonder if we cannot compress the columns of the matrix or, even better, change the definition of the matrix so they become the same.

\begin{figure}[t!]
\begin{center}
\begin{BVerbatim}
G ATTAGATACAT
A TTAGATACAT-
T TAGATACAT--
T AGATACAT---
A GATACAT----
G ATACAT-----
A TACAT------
T ACAT-------
A CAT--------
C AT---------
A T----------
T -----------
\end{BVerbatim}
\hspace{10ex}
\begin{BVerbatim}
T -----------
T ACAT-------
T AGATACAT---
C AT---------
G ATACAT-----
G ATTAGATACAT
A CAT--------
A GATACAT----
A T----------
A TACAT------
T TAGATACAT--
A TTAGATACAT-
\end{BVerbatim}
\hspace{20ex}
\begin{BVerbatim}
A TTAGATACAT
T TAGATACAT-
T AGATACAT--
A GATACAT---
G ATACAT----
A TACAT-----
T ACAT------
A CAT-------
C AT--------
A T---------
T ----------
- ----------
\end{BVerbatim}
\hspace{10ex}
\begin{BVerbatim}
T ----------
- ----------
T ACAT------
T AGATACAT--
C AT--------
G ATACAT----
A CAT-------
A GATACAT---
A T---------
A TACAT-----
T TAGATACAT-
A TTAGATACAT
\end{BVerbatim}
\end{center}
\caption{How we compute columns 0 and 1 in Figure~\ref{fig:suffixes} by sorting the characters in $S [0..n - 1] = \mathtt{GATTAGATACAT}$ and $S [1..n - 1]$ into the lexicographic order of the suffixes that immediately follow them.}
\label{fig:comparison}
\end{figure}

Suppose that, instead of considering the suffixes of $S$, we consider their cyclic shifts.  In case $S$ is periodic --- such as the string {\tt GATAGATA} from Section~\ref{sec:positional} --- we delimit it with a special character {\tt \$} smaller than all the characters in the alphabet.  This makes all the columns of the PBWT the same --- equal to the first column with a {\tt \$} inserted --- and can thus be stored in a total of $O (n \log \sigma)$ bits.  {\em This permutation of the characters in $S$ is the BWT of $S$.}  Figure~\ref{fig:BWT} shows how each column in the PBWT of $S$ would be computed with this change.

\begin{figure}[t!]
\begin{center}
\begin{BVerbatim}
G ATTAGATACAT$
A TTAGATACAT$G
T TAGATACAT$GA
T AGATACAT$GAT
A GATACAT$GATT
G ATACAT$GATTA
A TACAT$GATTAG
T ACAT$GATTAGA
A CAT$GATTAGAT
C AT$GATTAGATA
A T$GATTAGATAC
T $GATTAGATACA
$ GATTAGATACAT
\end{BVerbatim}
\hspace{10ex}
\begin{BVerbatim}
T $GATTAGATACA
T ACAT$GATTAGA
T AGATACAT$GAT
C AT$GATTAGATA
G ATACAT$GATTA
G ATTAGATACAT$
A CAT$GATTAGAT
A GATACAT$GATT
$ GATTAGATACAT
A T$GATTAGATAC
A TACAT$GATTAG
T TAGATACAT$GA
A TTAGATACAT$G
\end{BVerbatim}
\caption{How we compute all the columns of the PBWT --- which are equal to the BWT --- if we consider the cyclic shifts of $S = \mathtt{GATTAGATACAT}$ delimited by {\tt \$}, instead of the suffixes of $S$.}
\label{fig:BWT}
\end{center}
\end{figure}

We note as a historical aside that Bird and Mu~\cite{bird2004functional} used a similar idea in their paper on inverting the BWT in a functional setting, but ten years before the PBWT was published:
\begin{quotation}
``Then (4) states that the following transformation on the sorted rotations is the identity: move the last column to the front and re-sort the rows on the new first column. As we will see, this implies that the permutation that produces the first column from the last column is the same as
that which produces the second from the first, and so on.''
\end{quotation}
Interestingly, they were also motivated by providing greater insight into the BWT:
\begin{quotation}
``It often puzzles people, at least on a first encounter, as to why the BWT is
invertible and how the inversion is actually carried out.  We identify the fundamental
reason why inversion is possible and use it to derive the inverse transform from
its specification.''
\end{quotation}
Unfortunately, their paper had little impact on how bioinformaticians teach the BWT, perhaps because they focused on inversion instead of search or because understanding their paper really requires knowledge of Haskell.

We can now consider substring search using the BWT as a prefix search --- which is a special case of a positional search --- using a PBWT whose columns are all the same.  This may be easier for students to grasp, since usually the Last-to-First permution applied at each step during a search in a BWT is presented as an automorphism on one string, but now we can present it as a bijection from strings to other strings.  (As we briefly mention in Section~\ref{sec:conclusion}, this also simplifies some more advanced concepts.)

The last issue we should consider --- which Bird and Mu did not, since they were interested only in inversion and not search --- is how to sample permutations such that we can report the locations of the occurrences of $P$ that we find.  In Section~\ref{sec:positional} we sampled the permutation for every $\lceil \lg n \rceil$th column of the PBWT, but if we store the permutation for the BWT --- that is, the suffix array --- the we use $O (n \log n)$ bits, and if we do not store anything, then we cannot quickly report the locations of the occurrences of $P$.

A solution that ties in nicely with our presentation and the standard presentation of the FM-index is to switch from sampling columns to sampling diagonals.  All the characters in a diagonal in the matrix whose rows are the unsorted cyclic shifts, are the same character in the original string.  Therefore, if we sample those diagonals --- which is the same as sampling regularly in the original string --- then the sampled positions are the same in every column in the PBWT.  This is the standard suffix-array sample used in the FM-index and, with it, the space in bits is proportional to $n \lg n$ divided by distance between sampled positions in the original string, and the number of backward steps we take before finding the location of an occurrence of $P$ is still bounded by that distance.  Figure~\ref{fig:diagonals} shows the matrix for our example with the sampled positions highlighted in red, before and after we compute the BWT by sorting.

\begin{figure}[t!]
\begin{center}
\tt
\begin{tabular}{c}
\textcolor{red}{G}ATTA\textcolor{red}{G}ATAC\textcolor{red}{A}T\$\\
ATTA\textcolor{red}{G}ATAC\textcolor{red}{A}T\$\textcolor{red}{G}\\
TTA\textcolor{red}{G}ATAC\textcolor{red}{A}T\$\textcolor{red}{G}A\\
TA\textcolor{red}{G}ATAC\textcolor{red}{A}T\$\textcolor{red}{G}AT\\
A\textcolor{red}{G}ATAC\textcolor{red}{A}T\$\textcolor{red}{G}ATT\\
\textcolor{red}{G}ATAC\textcolor{red}{A}T\$\textcolor{red}{G}ATTA\\
ATAC\textcolor{red}{A}T\$\textcolor{red}{G}ATTA\textcolor{red}{G}\\
TAC\textcolor{red}{A}T\$\textcolor{red}{G}ATTA\textcolor{red}{G}A\\
AC\textcolor{red}{A}T\$\textcolor{red}{G}ATTA\textcolor{red}{G}AT\\
C\textcolor{red}{A}T\$\textcolor{red}{G}ATTA\textcolor{red}{G}ATA\\
\textcolor{red}{A}T\$\textcolor{red}{G}ATTA\textcolor{red}{G}ATAC\\
T\$\textcolor{red}{G}ATTA\textcolor{red}{G}ATAC\textcolor{red}{A}\\
\$\textcolor{red}{G}ATTA\textcolor{red}{G}ATAC\textcolor{red}{A}T
\end{tabular}
\hspace{10ex}
\begin{tabular}{c}
T \$\textcolor{red}{G}ATTA\textcolor{red}{G}ATAC\textcolor{red}{A}\\
T AC\textcolor{red}{A}T\$\textcolor{red}{G}ATTA\textcolor{red}{G}A\\
T A\textcolor{red}{G}ATAC\textcolor{red}{A}T\$\textcolor{red}{G}AT\\
C \textcolor{red}{A}T\$\textcolor{red}{G}ATTA\textcolor{red}{G}ATA\\
\textcolor{red}{G} ATAC\textcolor{red}{A}T\$\textcolor{red}{G}ATTA\\
\textcolor{red}{G} ATTA\textcolor{red}{G}ATAC\textcolor{red}{A}T\$\\
A C\textcolor{red}{A}T\$\textcolor{red}{G}ATTA\textcolor{red}{G}AT\\
A \textcolor{red}{G}ATAC\textcolor{red}{A}T\$\textcolor{red}{G}ATT\\
\$ \textcolor{red}{G}ATTA\textcolor{red}{G}ATAC\textcolor{red}{A}T\\
\textcolor{red}{A} T\$\textcolor{red}{G}ATTA\textcolor{red}{G}ATAC\\
A TAC\textcolor{red}{A}T\$\textcolor{red}{G}ATTA\textcolor{red}{G}\\
T TA\textcolor{red}{G}ATAC\textcolor{red}{A}T\$\textcolor{red}{G}A\\
A TTA\textcolor{red}{G}ATAC\textcolor{red}{A}T\$\textcolor{red}{G}
\end{tabular}
\caption{If we sample diagonals instead of columns, which is the same as sampling regularly in the original string, then the sampled positions are the same in every column of the PBWT.}
\label{fig:diagonals}
\end{center}
\end{figure}

\section{Conclusion}
\label{sec:conclusion}

We have shown that the PBWT and its support of positional search can be developed in a sequence of incremental steps --- each of which an undergraduate class should be able to figure out with a reasonable amount of guidance --- and that the BWT and the FM-index's support of substring search can be developed as positional search in the cyclic shifts of a string, whose PBWT has only one distinct column (which is the BWT).  Although this development is longer than simply presenting the BWT and FM-index directly, many students find the direct presentation hard to follow, and very few undergraduates could be expected to work out the steps for themselves --- often considered the best way to learn --- even with guidance.

Apart from teaching the BWT in a more accessible way, considering it as a refinement of the PBWT may have other advantages.  For example, in retrospect we think that working on the PBWT first would have made it easier to develop Nishimoto and Tabei's~\cite{nishimoto2021optimal} constant-time implementation of LF mapping with the r-index~\cite{gagie2020fully}.  This is because the LF map on the BWT is an automorphism but on the PBWT it is a bijection from strings to other strings, which can be sped up with a simple variation of fraction cascading~\cite[Section 4]{brown2021rlbwt}.

\bibliographystyle{plain}
\bibliography{PBWT2BWT}

\end{document}